\input{aipcheck}

\documentclass[
    ,final            
  ]
  {aipproc}

\layoutstyle{6x9}

\begin{document}

\title{A Constraint on brown dwarf formation via ejection: radial
  variation of the stellar and substellar mass function of the young
  open cluster IC~2391}

\classification{97.20.Vs}
\keywords {stars: low-mass, brown dwarfs --- stars: luminosity
  function, mass function --- open clusters and associations:
  individual (IC 2391)}

\author{S. Boudreault and C.A.L. Bailer-Jones}{
  address={Max-Planck-Institut f\"ur Astronomie, K\"onigstuhl 17,
  Heidelberg, GERMANY 69117}
}

\begin{abstract}
  Using the Wide Field Imager (WFI) at the ESO 2.2m telescope at La
  Silla and the CPAPIR camera at the CTIO 1.5m telescope at Cerro
  Tololo, we have performed an extensive, multiband photometric survey
  of the open cluster IC~2391 (D$\sim$146pc, age$\sim$50\,Myr, solar
  metallicity).  Here we present the results from our photometric
  survey and from a spectroscopic follow-up of the central part of the
  survey.
\end{abstract}

\maketitle


\section{SURVEY DETAILS AND CANDIDATE SELECTION PROCEDURE}

Our survey consists of 4 WFI fields (deep fields) in the central part
of the cluster observed in four medium band filters, namely 770/19,
815/20, 856/14 and 914/27, where the filter name notation is central
wavelength / FWHM in nm, and the broad band filter $R_{\rm c}$, 26 WFI
fields (radial fields, were chosen to extend predominantly in the
direction of constant Galactic latitude) with the broad bands $R_{\rm
  c}$ and $J$ and the bands 815/20 and 914/27, and 5 WFI fields
(outward fields) with $J$, 815/20 and 914/27, for a total coverage of
$\sim$10.9 sq.  deg.  The survey extends to a detection limit down to
0.02\,M$_\odot$ (10$\sigma$) and is centred on RA=08:35:45.1
DEC=-52:35:58.0.  Candidates were first selected based on
colour-magnitude diagrammes and a second selection was performed using
colour-colour diagrammes.  Third, astrometry was used to remove
objects with high proper motion.  Finally, non-candidates were
rejected based on a discrepancy between the observed magnitude in
815/20 and the magnitude in this band computed with the NextGen model
and our estimation of $T_{\rm eff}$.

\section{RESULTS : MASS FUNCTION OF THE DEEP AND OUTWARD FIELDS}

\begin{figure}
  \includegraphics[height=0.3\textheight]{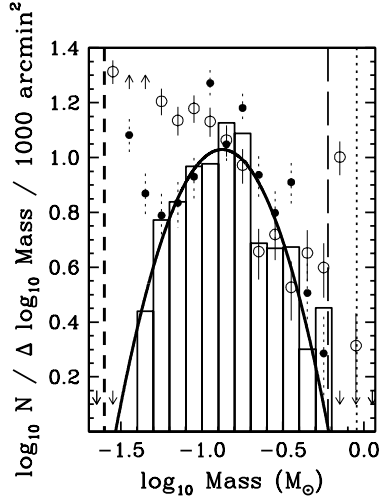}
  \caption{Filled dots represent the MF based on the four deep fields.
    Open dots represent the MF based on the outward fields. Also, we
    present again the 10$\sigma$ detection limit, the MF of all fields
    observed in $R_{\rm c}$, 815/20, 914/27 and $J$ within 2.1 from
    the cluster centre (histogram) and its log normal fit. The
    vertical thin dotted and thin dashed line lines are the mass for
    which saturation start to occur in the short exposures for outward
    and deep field respectively.\label{fig:mf-deep-89j-fit}}
\end{figure}

In both cases we see an apparent rise in the number of objects below
0.05\,M$_\odot$ (logM=-1.3, Fig. \ref{fig:mf-deep-89j-fit}, but we
concluded that this is an artefact of residual contamination by field
M dwarfs. This was also observed by \cite{barrado2004}. The fact that
we do not see this rise in the radial fields is because they were
observed with both the $J$ and $R_{\rm c}$ filters in addition to the
medium band filters. This provides a longer spectral baseline (better
determination of the energy distributions and helps rejection of field
M dwarfs based on observed magnitude vs. predicted magnitude from
models). Another apparent rise in the mass function (MF) over the
0.5--1.0\,M$_\odot$ interval (also observed for NGC~2547
\cite{jeffries2004}) is due to background giants. Red giant
contamination may be reduced by using medium bands such as 770/19,
815/20, 856/14 and 914/27, and theoretical colours of red giants
\cite{hauschildt1999b}. Our spectroscopic follow-up has confirmed that
selection based on these filters resulted in no red giant contaminants
among a sample.

\section{RESULTS : RADIAL VARIATION OF THE STELLAR AND SUBSTELLAR
  POPULATION}

\begin{figure}
  \includegraphics[height=0.3\textheight]{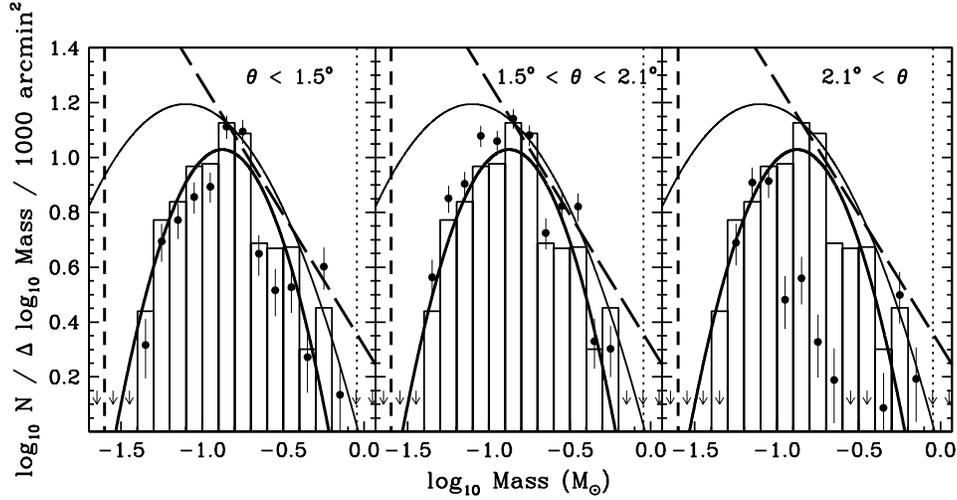}
  \caption{MF based on photometry for all radial fields. The
    10$\sigma$ detection limit is shown as a vertical dashed line. The
    MF fit for IC~2391 from \cite{barrado2004} and for the galactic
    field \cite{chabrier2003} stars are shown as thick dashed thin
    solid lines respectively. The thick solid line is the fitted
    lognormal function of the MF of IC~2391 (within 2.1). Dots in each
    panel represent the MF of (left) fields within 1.5 of the cluster
    centre, fields within the annulus from 1.5 to 2.1 and (right) the
    MF of fields outside of 2.1. Error bars are Poissonian arising
    from the number of objects observed in each bin. The histogram is
    the MF for all fields within 2.1 of the cluster centre. The
    vertical thin dotted line is the mass for which saturation start
    to occur in the short exposures.\label{fig:mf-r89j}}
\end{figure}

Radial variation is observed in the MF from 0.15 to 0.5\,M$_\odot$
(Fig. \ref{fig:mf-r89j}) and we argue that this is a signature of mass
segregation, presumably via dynamical evolution. This is consistent
with theoretical predictions since the age of IC~2391 is half of its
relaxation time (estimated at $\sim$105\,Myr). We do not observe a
significant radial variation in the MF bellow 0.15\,M$_\odot$.
Although this absence of radial variation of the brown dwarf (BD)
population would be in agreement with the ejection scenario of BD
formation, the fact that we do not observe a discontinuity in the MF
across the stellar/substellar boundary (0.072\,M$_\odot$) implies that
the ejection formation scenario is not a significant BD formation
mechanism in this cluster if this formation mechanism results in a
higher distribution velocity of BDs compared to stars
\cite{kroupa2003}. On the other hand, if ejection mechanism is the
unique BD formation path, then both BDs and stars should have the same
velocity dispersion \cite{bate2003}.

\section{RESULTS : SPECTROSCOPIC FOLLOW-UP}

\begin{figure}
  \includegraphics[width=1.0\textwidth]{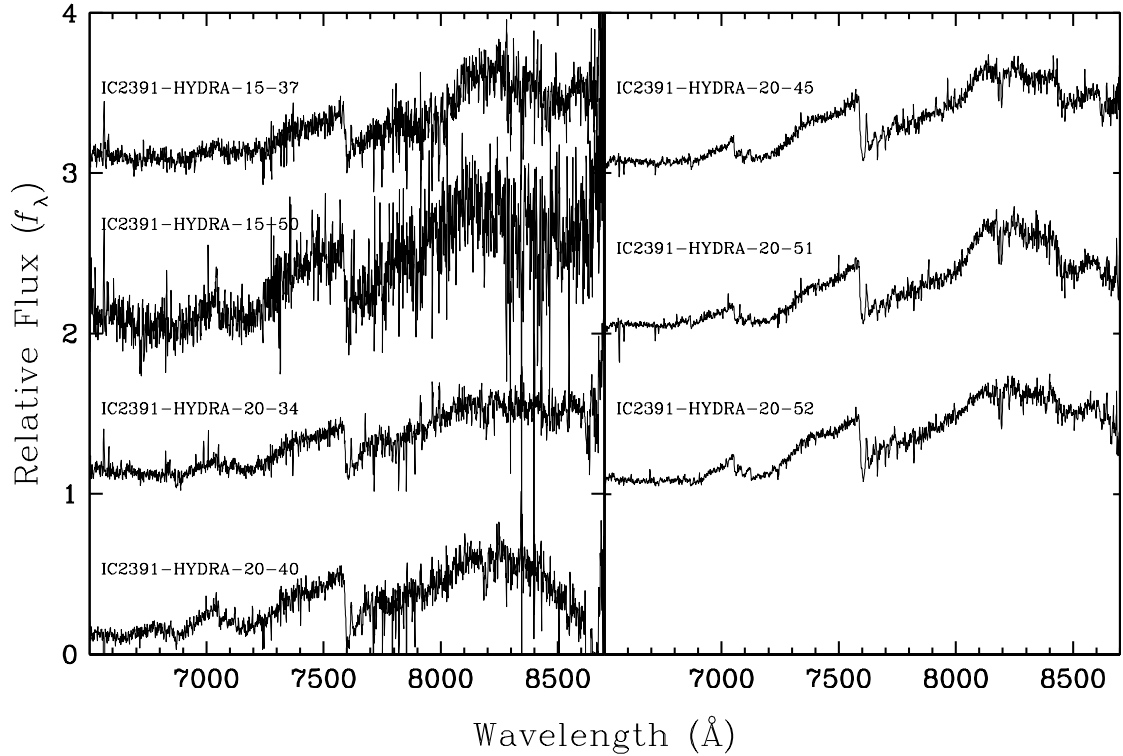}
  \caption{Spectra of the seven newly discovered BD members of the
    IC~2391 cluster found in our survey. Objects are given the
    notation IC~2391-HYDRA-ZZ-YY where ZZ is the field number and YY a
    serial identification number (ID). \label{fig:ic2391-bd}}
\end{figure}

From our preliminary spectroscopic follow up, all 77 spectra analysed
are M-dwarfs (field or in the cluster), demonstrating the efficiency
of our method to avoid red giant contaminants in our photometric
selection using medium filter. We also observed that H$\alpha$
emission line cannot be used as a membership criterion in IC~2391 when
taking fibre spectroscopy because of background contamination, if
background is assumed to be uniform. About 31\% of our photometric
candidates are true physical members of the cluster, where 7 are new
spectroscopically-confirmed BD members of IC~2391 (Fig.
\ref{fig:ic2391-bd}).

\section{CONCLUSIONS}

We have performed a photometric survey of the open cluster IC~2391 to
study the radial dependence of the MF. We have 1734 photometric
candidates for the outward fields, 499 from the deep fields and 954
from the radial fields, which gives $\sim$8 candidates per 100 arcmin.
Ejection formation scenario is not a significant BD formation
mechanism if it results in a higher velocity dispersion of BDs
compared to stars. However, if ejection mechanism is the unique BD
formation path, then both BDs and stars should have the same velocity
dispersion. Variations in the colours of the main (field star) locus
in the CMDs are due to spatial extinction-induced variations in
background star contaminants. Selection based on medium filters (such
as 770/19, 815/20, 856/14 and 914/27) resulted in no red giant
contaminants while a broad spectral baseline (such as the use of
$R_{\rm c}$ to $J$) was successful in reducing M-dwarf
background/foreground contaminants. About 31\% of our candidates from
our spectroscopic follow-up are physical members, from which 7 are
BDs.  H$\alpha$ emission line cannot be used as a membership criterion
in IC~2391 when taking fibre spectroscopy if background is assumed to
be uniform (we recommend background subtraction to be performed in a
similar way as the one done by \cite{carpenter1997}, where the same
fibers for the science targets were also used for sky subtraction but
shifted 6 arcsec away). More results and informations are presented in
\cite{boudreault2008}.







\bibliographystyle{aipproc}   




\end{document}